# LCLS-II Technical Note

Untrapped HOM Radiation Absorption in the LCLS-II Cryomodules

LCLS-II TN-14-08

11/4/2014

K. Bane, C. Nantista, C. Adolphsen, T. Raubenheimer,
SLAC, Menlo Park, CA 94025, USA
A. Saini, N. Solyak, V. Yakovlev,
FNAL, Batavia, IL 60510, USA

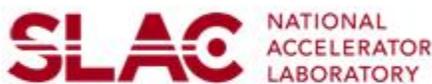
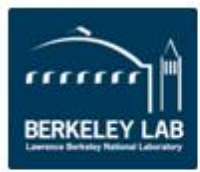
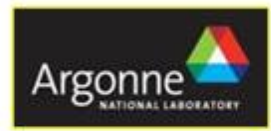
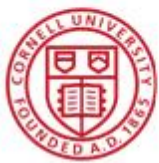
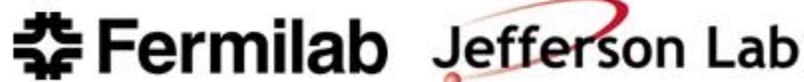

# UNTRAPPED HOM RADIATION ABSORPTION IN THE LCLS-II CRYOMODULES*

K. Bane, C. Nantista[#], C. Adolphsen and T. Raubenheimer, SLAC, Menlo Park, CA 94025, USA,
A. Saini, N. Solyak and V. Yakovlev, FNAL, Batavia, IL 60510, USA

*Abstract*

The superconducting cavities in the continuous wave (CW) linacs of LCLS-II are designed to operate at 2 K, where cooling costs are very expensive. One source of heat is presented by the higher order mode (HOM) power deposited by the beam. Due to the very short bunch length−especially in L3 the final linac−the LCLS-II beam spectrum extends into the terahertz range. Ceramic absorbers, at 70 K and located between cryomodules, are meant to absorb much of this power. In this report we perform two kinds of calculations to estimate the effectiveness of the absorbers and the amount of beam power that needs to be removed at 2 K.

## INTRODUCTION

While the use of superconducting accelerating cavities in large particle accelerator facilities offers many advantages in areas such as RF efficiency and feasible beam parameter ranges, a major expense of operating such a machine is the power required of the cryogenic plant. Care must be taken to minimize both static and dynamic heat loads. One element of the latter, particularly relevant in a high-current, short bunch, CW facility like LCLS-II, is higher-order-mode (HOM) electromagnetic field power generated by the beam in passing through the cavities and beamline elements.

LCLS-II will run with a CW megahertz bunch train of initial current 62 μA, but eventually upgradable to 0.3 mA. HOM heat load is of particular concern in the 20 cryomodules of the L3 linac region, after the second bunch compressor, where the rms length of the 300 pC bunches will be only $\sigma_z$ = 25 μm. The main generators of HOM power are the 35 mm radius irises of the nine-cell periodic L-band accelerator cavities, though other features, such as inter-cavity bellows and beam pipe radius transitions between regions, play a role. For a quantitative treatment of generated HOM (or more accurately wakefield) power in LCLS-II see [1,2].

HOM coupler ports incorporated in the end pipes of the cavities provide damping against build-up of higher-order cavity mode fields. Since the spectrum of excited HOM power extends well beyond the beam pipe cutoff, annular ceramic RF absorbers are included in the drifts between the 8-cavity cryomodules in the hope of absorbing much of this untrapped wakefield radiation.

In what follows, we describe and compare two attempts to theoretically characterize the relative HOM power lost in the different cryogenic environments (2 K, 70 K) by assessing the effectiveness of the HOM absorbers. We would like as much of the power as possible to be lost in the 70 K absorbers rather than in the NC beam pipes and bellows between the cavities, for which heat is removed by the 2 K cooling system. The first method uses a numerical, S-matrix approach, and the second involves an analytical diffusion-type calculation. This topic was previously addressed for the European XFEL project using a ray tracing method and the diffusion approach presented here [3]. Also, the S-matrix calculation has been previously applied to the ILC cryomodules [4]. We focus here on the maximum average beam current.

## S-MATRIX APPROACH

Most of the LCLS-II linac can be seen as a periodically repeated sequence of cryomodule elements joined at their 39 mm-radius beam pipe ports. This periodic unit consists of eight nine-cell ILC-type cavities, each followed by a bellows, a long beam pipe drift (through the quad), an absorber and a shorter drift section. The absorber, a suspended ring of ceramic recessed in a pillbox diameter step, has an integrated bellows, which we treat as part of the same element.

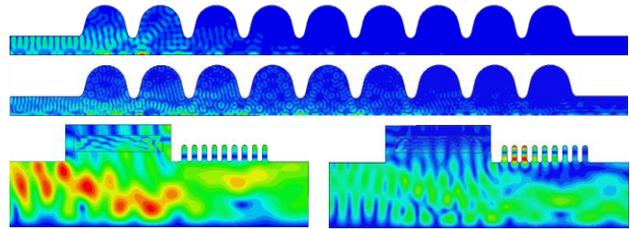

Figure 1: Geometries and field plots from HFSS S-matrix calculations: cavity and absorber at 12 & 20 GHz.

At a number of discrete frequencies, 4, 8, 12, 16, 20 and 40 GHz, we used the field solver HFSS [5] to calculate the scattering matrix for each element (cavity, bellows, drifts and absorber) for all $TM_{0n}$ monopole modes propagating in the beam pipe at each respective frequency. The niobium cavities were modelled as perfect conductors. The bellows and drift pipes were assumed to be copper (or copper plated), for which an electrical conductivity in this temperature and frequency regime, ranging from 1.97–0.914×10⁹/Ωm between 4–40 GHz, was calculated according to

$$\sigma_{Cu,eff}(\omega) = (\omega Z_0/2c)\text{Re}(\mathcal{Z}_{Cu,eff}(\omega))^{-2}, \quad (1)$$

___________________________________________
*Work supported by the U.S. Department of Energy under contract DE-AC02-76SF00515.
[#] nantista@slac.stanford.edu

where $Z_0$ is the impedance of free space and $\mathcal{Z}_{Cu,eff}$ is the surface impedance from Eq. (6) below for the extreme anomalous skin effect [6]. We also considered the case where the bellows and drifts are stainless steel, for which we took $\sigma_{ss} = 1.85\times10^6/\Omega$m. For the absorber element, only the ceramic material was given loss, to represent heat removed. We assumed here Re$(\varepsilon)/\varepsilon_0 = 15$ and tan $\delta = 0.18$, estimated from available data [7] out to 40 GHz. Fig. 1 shows geometries and sample field plots.

Armed with these matrices for a set of $M$ sequential elements, we can specify a set of equations relating the left ($l$) and right-going ($r$) waves of $N$ propagating modes at the junctions as follows:

$(l_{n,1}, l_{n,2},..., l_{n,M}, r_{n+1,1}, r_{n+1,2},..., r_{n+1,M})^T =$
$\mathbf{S}_n (r_{n,1}, r_{n,2}, ..., r_{n,M}, l_{n+1,1}, l_{n+1,2},..., l_{n+1,M})^T$, (2)

With a constant driving vector $d_{n,1} = e^{-i\omega z_n/c}$ after each cavity ($z_n$ being its downstream position) and 0 elsewhere, representing the wakes of a speed-of-light bunch, and speed-of-light phased periodic boundary conditions imposed at the ends of the sequence, determining the junction fields amounts to solving a matrix equation of the form $\mathbf{A}y = d$, where $\mathbf{A}$ is of order $2M(N+1)$.

Now, correcting for the generated waves, we can solve for the relative power dissipated in each element as the difference between incoming and outgoing power, summed over the modes. That is,

$$p_n = \Sigma_{m=1,N} (|r_{n,m}|^2 + |l_{n+1,m}|^2 - |l_{n,m}|^2 - |r_{n+1,m}|^2) + |r_{n+1,1}|^2 - |r_{n+1,1} - d_{n+1,1}|^2. \quad (3)$$

This technique is described in more detail in [4]. The ratio of the $p$ for the absorber element to the sum of all $p$'s gives us a measure of the effectiveness of the absorber in reducing the losses at 2 K. As the number of modes for each case was the frequency in gigahertz divided by four, this approach is limited from going much higher than about 40 GHz by the S-matrix size.

## DIFFUSION APPROACH

Another way to estimate the distribution of HOM power absorption is to use a diffusion-like model [3], in which the radiation fills the available volume like a gas. This is a good approximation well above cutoff and where the surface reflection coefficient is close to unity. The beam line elements can be grouped by type, for each of which, the power absorption can be characterized by

$$I_i^{abs} = n_i S_i \int \frac{dP_0}{d\omega}(\omega) \text{Re}(\mathcal{Z}_i(\omega)) d\omega; \quad (4)$$

where $i$ denotes a particular type of element (RF cavity, bellows, end pipe or absorber), $n_i$ is the element quantity, $S_i$ its surface area, $\mathcal{Z}_i$ its surface impedance and $dP_0/d\omega$ the HOM spectral density obtained using:

$$dP_0(\omega) = q_b^2 f_r \text{Re}[Z_w(\omega)] e^{-(\omega\sigma_z/c)^2} d\omega/2\pi, \quad (5)$$

where $Z_w$ is the cryomodule impedance. To obtain the spectrum, we begin with the point charge wake of a TESLA cryomodule: $W(s) = 344 \exp(-\sqrt{s/s_0})$ V/pC, with $s_0 = 1.74$ mm [7]. We then Fourier transform this to obtain $Z_w(\omega)$.

The resulting spectrum $P_0(f)$ (with $f = \omega/2\pi$) — i.e. the total power integrated from $f$ to 2 THz — is shown in Fig. 2 for the L3 linac parameters. Note that it is the derivative of this function divided by $2\pi$ that gives $dP_0/d\omega$ of Eq. (4). With an rms bunch length of $\sigma_z = 25$ μm, a bunch charge $q_b = 300$ pC and a beam of repetition rate $f_r = 1$ MHz, the total steady state power ($P_\text{HOM}$) deposited in a cryomodule is 13.8 W [1, 2]. For comparison, the main RF cryomodule heat load is approximately 100 W.

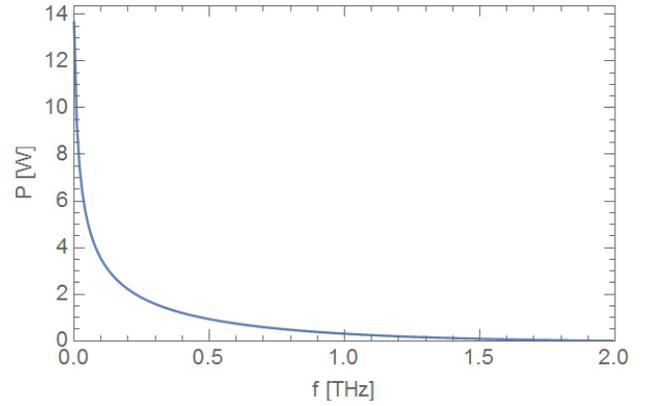

Figure 2: Integrated steady state wakefield power from $f$ up to 2 THz generated by 300 pC 25 μm bunches at 1 MHz in L3. Of the 13.8 W total lost by beam, about 50% is lost above 20 GHz, and 25% above 100 GHz.

Here too, two cases were considered, in which the bellows and beam pipe surfaces were assumed to be either copper or stainless steel. For frequency > 1 GHz at 2 K, copper exhibits the extreme anomalous skin effect. Its effective impedance is expressed as [6]

$$\mathcal{Z}_{Cu,eff}(\omega) = A\omega^{2/3}(1+i\sqrt{3}), \quad (6)$$

where $A = 3.3\times10^{-10}$ ($\Omega$ s$^{2/3}$) is a material constant independent of RRR. The surface impedance of stainless steel is given by:

$$\mathcal{Z}_{SS}(\omega) = (1+i)\sqrt{\omega Z_0/(2c\sigma_{ss})}, \quad (7)$$

where $\sigma_{ss}$ is the electrical conductivity of stainless steel, for which we here used $10^6/\Omega$m. As before, the power absorbed in the superconducting niobium cavities was taken to be negligible.

To estimate the surface impedance of the absorber material we use the following:

$$\alpha(\omega) = (\omega \tan\delta/2c)\sqrt{\text{Re}(\varepsilon)/\varepsilon_0} \quad (8)$$
$$\tan\delta \equiv \text{Im}(\varepsilon)/\text{Re}(\varepsilon), \quad \delta_s = 1/\alpha, \quad \sigma = \omega\text{Im}(\varepsilon);$$

where $\alpha$ is the attenuation coefficient in the dielectric medium, $\varepsilon$ its permittivity, tan $\delta$ the loss tangent, and $\delta_s$ the field penetration skin depth. (Here values of 16.5 and

0.2 were used respectively for Re($\varepsilon$)/$\varepsilon_0$ and tan $\delta$.) The surface impedance of the absorber is given by:

$$\mathcal{Z}_{Absorber}(\omega) = 1/(\sigma\delta_s) \quad (9)$$

As for the $n_iS_i$ area factors, the bellows and pipe total was estimated to exceed the ceramic by a factor of 57. Inserting the required parameters in Eq. (4) calculated from Eqs. (5), (6), (7) and (9), one can estimate the $I_i^{abs}$ for each element type and hence the power absorption for each in an L3 cryomodule from:

$$P_i = P_{HOM}\left(I_i^{abs}\bigg/\sum_{j=1}^{3} I_j^{abs}\right) \quad (10)$$

## RESULTS

Each of these methods yields an estimate for the fractional heating distribution among beamline elements of the high-frequency wakefield power generated by the LCLS-II beam. Table 1 tabulates side-by-side results at given frequencies for the fraction that goes into the 2 K cryogenics, i.e. is not extracted to 70 K by the beamline HOM absorber at the end of each cryomodule. The numbers are similar. The final row shows, for the diffusion model, the total integrated percent, weighted by the spectral density function, the derivative of Fig. 2. It suggests the absorber is considerably less effective at higher frequency.

Table 1: Percent of Untrapped HOM Power to 2 K.

| f (GHz) | Copper | | Stainless Steel | |
|---|---|---|---|---|
| | S-matrix | diffusion | S-matrix | diffusion |
| 4 | 0.75 | 0.35 | 19 | 13 |
| 8 | 3.5 | 0.55 | 49 | 18 |
| 12 | 0.5 | 0.70 | 23 | 21 |
| 16 | 0.55 | 0.85 | 10 | 24 |
| 20 | 1.1 | 1.0 | 47 | 26 |
| 40 | 1.1 | 1.6 | 35 | 33 |
| Total | | 2.7 | | 39 |

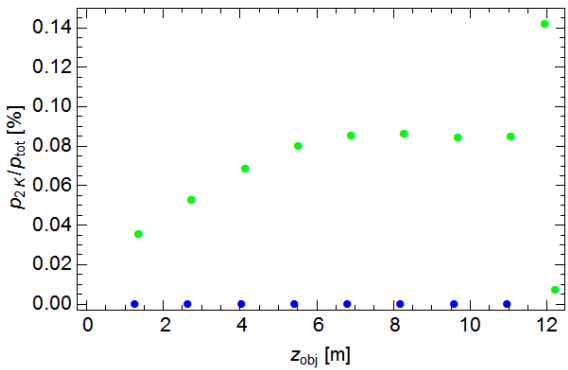

Figure 3: Relative power distribution at 4 GHz in the elements along the cryomodule. Blue dots are the cavities. The absorber, at > 99% at 12 m, is offscale.

Stainless steel does not appear to be an acceptable beamline surface material for this machine, based on both the higher power absorption and consideration of thermal conduction to the liquid helium bath. Bellows and connecting drift pipes should be copper-plated.

A sample plot is shown in Fig. 3 of the element power distribution from the S-matrix technique at 4 GHz. As this depends on mode phase lengths, we estimated the behaviour away from our discreet S-matrix frequencies by systematically extending cavity end pipes, analytically. Fig. 4 plots the effect on the 2 K power at 20 GHz. Statistical results over 2,000 steps are given in Table 2.

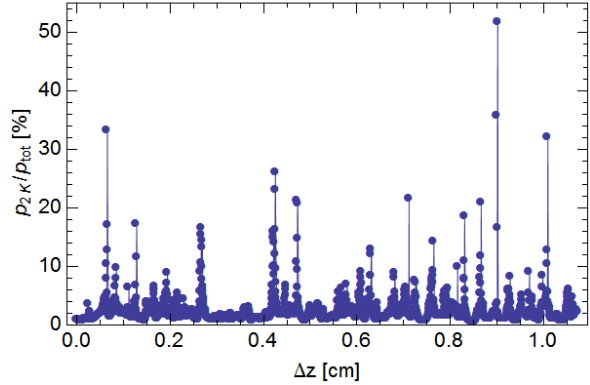

Figure 4: Percentage of power deposited at 2 K, $p_{2K}/p_{tot}$, vs. added cavity pipe length at $f$ = 20 GHz.

Table 2: Element Spacing Statistics of Power to 2 K (Cu).

| f (GHz) | average | rms | 9th decile |
|---|---|---|---|
| 4 | 2.3 | 6.5 | 3.1 |
| 8 | 1.2 | 4.2 | 1.7 |
| 12 | 1.3 | 4.6 | 1.4 |
| 16 | 1.5 | 4.3 | 2.4 |
| 20 | 5.3 | 8.5 | 11.5 |
| 40 | 2.5 | 2.8 | 4.3 |

Since driving only the lowest mode ($d_{n,1}$) is not what the beam actually does, we tested the sensitivity of our results to this assumption. At $f$ = 16 GHz, we repeated the power vs. cavity pipe length calculation driving in turn each of the other three propagating monopole modes. The averages of $p_{2K}/p_{tot}$ resulting from driving modes 1–4, respectively, were 1.5%, 1.5%, 1.4% and 1.7%. It seems mixing makes the calculation fairly insensitive to this drive detail.

## CONCLUSION

Two complementary approaches applied to the LCLS-II linac have helped to characterize the distribution of heating due to monopole wakefield power generated by the tightly-bunched CW beam. They provide us some confidence in the effectiveness of the beamline HOM absorbers, suggesting that no more than a few percent of this power will present added load to the 2 K cryogenics system. We should note, however, that the absorber material, for which we assume constant parameters, has not been characterized above 40 GHz.

## ACKNOWLEDGMENT

The authors wish to thank Martin Dohlus for visiting us to explain the European XFEL effort on the topic of this report.